\documentclass[conference]{IEEEtran}
\IEEEoverridecommandlockouts
\usepackage{cite}
\usepackage{amsmath,amssymb,amsfonts}
\usepackage{algorithmic}
\usepackage[ruled,vlined,linesnumbered]{algorithm2e}
\SetKwInput{KwIn}{Input}
\SetKwInput{KwOut}{Output}
\SetKwProg{Fn}{Function}{:}{}
\SetKw{Return}{Return}
\usepackage{graphicx}
\usepackage{textcomp}
\usepackage[dvipsnames]{xcolor}
\usepackage{subcaption}
\usepackage{booktabs}
\usepackage{multirow}
\usepackage{placeins}
 
\def\BibTeX{{\rm B\kern-.05em{\sc i\kern-.025em b}\kern-.08em
    T\kern-.1667em\lower.7ex\hbox{E}\kern-.125emX}}
\begin{document}

\title{Communication-Aware Placement and Pruning for Efficient Mixture-of-Experts Inference}

\author{
\IEEEauthorblockN{Xiao Shi, Yingying Sun, Jiangsu Du, Zhiguang Chen, Yutong Lu}
\IEEEauthorblockA{
Sun Yat-sen University\\
Guangzhou, China\\
Email: \{shix36, sunyy57\}@mail2.sysu.edu.cn, \{dujiangsu, chenzhg29, luyutong\}@mail.sysu.edu.cn  
}
}
\maketitle

\begin{abstract}

As MoE models scale to hundreds of experts, placement and pruning decisions increasingly dictate communication volume, affecting the performance of distributed inference across GPUs and nodes.
We propose CAP (Communication-Aware Assignment and Pruning), a framework that considers computation, communication and accuracy together for efficient MoE inference through expert placement and pruning.
It consists of three components:
(1) Co-activation driven expert placement, which groups frequently co-activated experts to reduce inter-device and inter-node communication;
(2) Communication-computation trade-off adjustment, which generates placements with different computational load and communication volume; and
(3) Communication-aware expert pruning, which selectively removes routing destinations to reduce communication with limited accuracy degradation.
By combining these components, CAP selects an efficient operating strategy for different hardware configurations. Across our single-node and multi-node experiments, it achieves 1.23$\times$--1.86$\times$ throughput improvement over DeepSeek EPLB and sequential placement in vLLM, and preserves better model accuracy at the same target speedup under lossy acceleration.

\end{abstract}


\section{Introduction}
 
The sparsely activated Mixture-of-Experts (MoE) architecture is increasingly used to scale the size of large language model (LLM) and boost performance~\cite{dai2024DeepSeekmoe,jiang2024mixtral,zhu2024llama,brown2020languagemodelsfewshotlearners,vaswani2023attentionneed,touvron2023llama}.
A key challenge in MoE inference with expert parallelism is the imbalance in workload distribution across devices.
Many efforts are devoted to adjusting expert placement and pruning for better performance.
However, as the number of experts increases, existing approaches generally overlook a critical factor: the fact that expert placement and pruning significantly affect communication volume, leading to sub-optimal performance.

\begin{figure}[!t]
    \centering
    \includegraphics[width=0.5\textwidth]{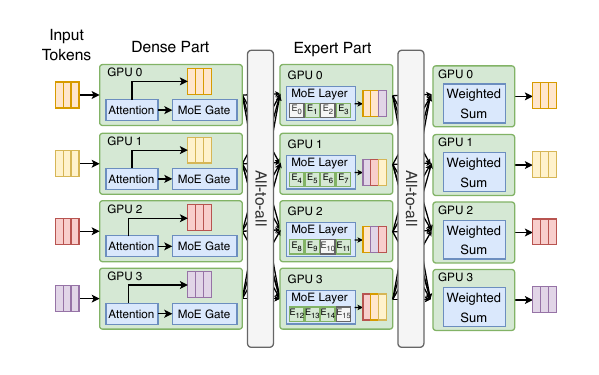}
    \vspace{-20pt}
    \caption{Expert parallelism combined with data parallelism}
    \vspace{-20pt}
    \label{fig:ep}
    
\end{figure}

MoE inference systems typically adopt a combination of expert and data parallelism.
As shown in Figure~\ref{fig:ep}, the data parallelism is applied to the dense attention part, where each device replicates the full weights of the dense layers, while expert parallelism is applied to the expert parts, where experts are partitioned by expert and distributed across all devices.
When input sequences arrive, they are processed in different devices separately, and then each token is routed to the devices hosting its target experts through a global all-to-all communication.
The token is then returned to the originating device through another global all-to-all communication, so that subsequent attention computation of sequences can be performed.

Expert placement and expert pruning have become common approaches for optimizing MoE inference systems.
Expert placement methods\cite{han2025gracemoegroupingreplicationlocalityaware,huang2024toward,yu2024moesysdistributedefficientmixtureofexperts,li2024acceleratingdistributedmoetraining,rajbhandari2022deepspeedmoeadvancingmixtureofexpertsinference,Nie_2023} are lossless.
Since the activation frequencies of different experts remain uneven, the distributed MoE inference generally suffers from severe load imbalance.
By co-locating frequently and infrequently activated experts on the same device, it can balance the computation load and improve overall inference efficiency.
Expert pruning methods~\cite{koishekenov2023memoryefficientnllb200languagespecificexpert,gupta2024lynxenablingefficientmoe,lu2024expertsequalefficientexpert,chen2022taskspecificexpertpruningsparse} are typically lossy.
They dynamically remove negligible experts for each token and trade a small amount of accuracy for higher inference speed.

Since prior MoE models featured only a small number of experts, existing approaches overlooked their impacts on communication volume and did not incorporate communication into their design.
For instance, Mixtral 8x7b~\cite{jiang2024mixtral}, released in 2023, has only 8 experts with a total of 46B parameters.
When distributed across eight devices, each device hosts exactly one expert, leaving no degree of freedom for expert placement to impact communication volume.
In contrast, newer MoE architectures are substantially more complex. The 30B version of Qwen3~\cite{yang2025qwen3}, released in 2025, includes 128 small experts, leaving substantial room for communication optimization through placement.
Similarly, existing expert pruning approaches often fail to deliver the expected performance benefits on modern MoE models, as considering computation alone without communication does not alleviate the dominant bottleneck.

In this paper, we propose CAP (Communication-Aware Assignment and Pruning), a framework that considers computation, communication and accuracy together for efficient parallel MoE inference.
The framework includes three components and integrates them in a unified pipeline for determining the optimal placement and pruning strategy.
First, we propose a co-activation driven expert placement approach that identifies frequently co-activated experts and generates a communication-oriented initial placement.
Second, we introduce an adjustment approach that starts from the communication-optimal placement and, based on runtime load balance statistics, generates a spectrum of candidate placements with different trade-offs between communication and load balance.
Third, we design the communication-aware expert pruning approach, which incorporates communication cost into the pruning decisions, enabling the system to achieve the desired speedup while minimizing the impact on accuracy.
By integrating these three components, CAP improves throughput across both single-node 8-GPU and multi-node settings compared with load-balance-oriented placement and default sequential placement, and preserves better model accuracy at the same target speedup under lossy acceleration.

\section{Background}
\subsection{Mixture-of-Experts Architecture}

Large Language Models (LLMs) have achieved remarkable gains as model scale increases, a phenomenon often described by the scaling law~\cite{devlin2019bert}. This trend has driven rapid growth in parameter counts, but also led to substantially increased training and inference costs. To mitigate these costs, GShard~\cite{lepikhin2020gshard} and other works~\cite{outrage} introduced the Mixture-of-Experts (MoE) architecture, which activates only a small subset of model parameters for each token. This design allows MoE models to maintain high capacity while significantly reducing per-token computation, and has since been adopted by state-of-the-art models such as DeepSeek-V3~\cite{liu2024DeepSeek}, GPT-4~\cite{achiam2023gpt}, and Qwen3~\cite{yang2025qwen3}.

In MoE LLMs, the dense FFN in each transformer layer is replaced by a set of experts. After the attention module, a gating network selects the top-$k$ experts for each token, where $k$ is typically small (e.g., $k=1$, $2$, or $8$). The output of the MoE layer is then computed as the weighted sum of the selected experts:
\begin{equation}
\text{MoE}(x) = \sum_{i=1}^{k} Gate_i(x) \cdot \text{Expert}_i(x)
\label{eq:moe}
\end{equation}
, where $Gate_i(x)$ denotes the normalized routing weight of the $i$-th selected expert for token $x$, and $\text{Expert}_i(x)$ denotes the corresponding expert output. In this way, only a sparse subset of experts participates in the computation of each token.

\subsection{Expert Parallelism in MoE Inference}

Because modern MoE models often contain a large number of parameters, the model usually cannot fit on a single device. To utilize the memory and computational resources of multiple devices, expert parallelism~\cite{lepikhin2020gshard} is widely adopted. In this paradigm, experts are distributed across devices, while dense components such as attention and projection layers are replicated. 

During inference, each device first processes different requests through the dense part of the model, and then uses the gating network to determine the experts required by each token. Since the selected experts may reside on different devices, tokens must be routed to the corresponding devices for expert computation. After expert computation is completed on the destination devices, the results are returned and aggregated to produce the final MoE output. As a result, expert routing introduces substantial communication overhead in distributed MoE inference.

This parallel execution strategy enables MoE models to scale beyond the memory capacity of a single device while distributing computation across multiple devices. However, it also makes inference performance sensitive to the routing distribution of tokens and the communication pattern induced by expert parallelism.

\begin{figure}
    \centering
    \includegraphics[width=0.95\linewidth]{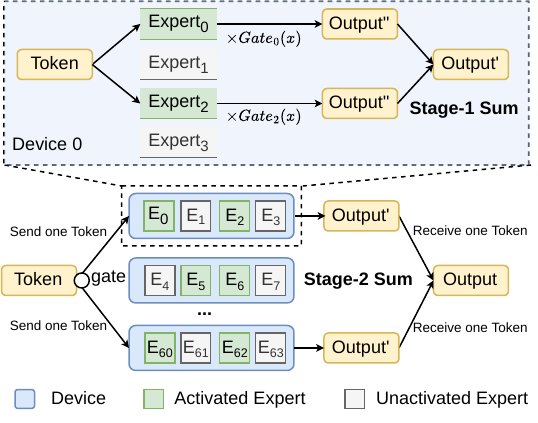}
    
    \caption{Zero-redundancy Token Transfer}
    \label{fig:zero-redundancy}
    \vspace{-20pt}
\end{figure}

\subsection{Communication in Expert Parallelism}
\label{sec:bg-comm}
In expert parallel inference, experts are distributed across multiple devices, and data dependencies are resolved through an all-to-all communication operation that routes tokens to their assigned experts.
In particular, as shown in Fig.~\ref{fig:zero-redundancy}, a token is routed to multiple experts and some experts may reside on the same device (e.g., expert 0 and expert 2 are colocated on device 0).
A naive implementation would send the token to each assigned expert (e.g., send to device0 twice), which is an expert-centric abstraction that overlooks device-level communication efficiency.
In contrast, existing systems typically adopt a device-centric design, where each token is sent once to each device that hosts activated experts, and all experts on the same device share the same copy.
After the experts produce their outputs, the outputs on the same device are locally aggregated via a weighted summation, and the aggregated result is sent for final combination.
Therefore, for a single token, the communication volume of the all-to-all communication operation primarily depends on the number of devices it needs to visit, rather than the number of experts it activates.

\subsection{Dynamic Expert Pruning }

Dynamic expert pruning is a widely used technique for improving MoE inference efficiency.
During inference, experts are selected based on their routing scores \(Gate_i(x)\), while those with low scores are ignored, as their contributions to the final output are negligible but incur additional computation overhead.
Existing dynamic methods typically retain a minimal subset of experts such that the aggregated routing score of the selected experts exceeds a predefined threshold.
However, existing methods typically overlook how pruning decisions affect communication overhead in distributed MoE inference. 
In particular, pruning not only impacts computation and model accuracy, but also alters the set of routing destinations visited by each token. 
As a result, different pruning decisions can lead to significantly different communication behaviors, even when their impact on accuracy is similar.
Pruning strategies should explicitly consider communication efficiency.

\subsection{Challenges}

Expert parallelism introduces several systems challenges, most notably load imbalance~\cite{hwang2023tuteladaptivemixtureofexpertsscale, gale2022megablocksefficientsparsetraining}. In MoE models, the numbers of tokens processed by different experts are often highly uneven. This results in stalls in all-to-all communication, where faster devices must wait for slower ones, degrading overall system efficiency.

Prior work mainly addresses this issue by adjusting expert placement across devices. Some approaches monitor expert popularity at runtime and co-locate hot and cold experts to improve load balance~\cite{he2022fastermoe, smartmoe}. However, these methods primarily treat expert placement as a load-balancing problem and do not explicitly consider its impact on communication volume. As a result, they may overlook an important source of overhead in MoE inference systems, where communication can become a major bottleneck.

Moreover, the relative importance of communication overhead and computation balance can vary substantially across hardware platforms. As a result, an optimization strategy that works well on one machine may be suboptimal on another. Therefore, MoE inference optimization should adapt to the communication and computation characteristics of the underlying hardware.

\section{Overview}

CAP consists of three components. First, we introduce a co-activation-driven expert placement method to reduce communication by grouping experts that tend to be activated together onto the same device. Second, we present a comm.-comp. trade-off adjustment method, which explores different placements with different trade-offs between communication cost and load balance, and selects the most suitable one according to hardware characteristics. Third, we introduce communication-aware expert pruning, which further reduces communication under a predefined accuracy threshold  by pruning routing destinations with low contribution and high communication cost. These three components are executed in an ordered pipeline, as later described in Section \ref{sec:overall-pipeline}.

\section{Co-activation Driven Expert Placement}
\label{sec:ap-1}
\subsection{Motivation}

\begin{figure}[t]
    \centering
    \begin{subfigure}[t]{0.48\linewidth}
        \centering
        \includegraphics[width=\linewidth]{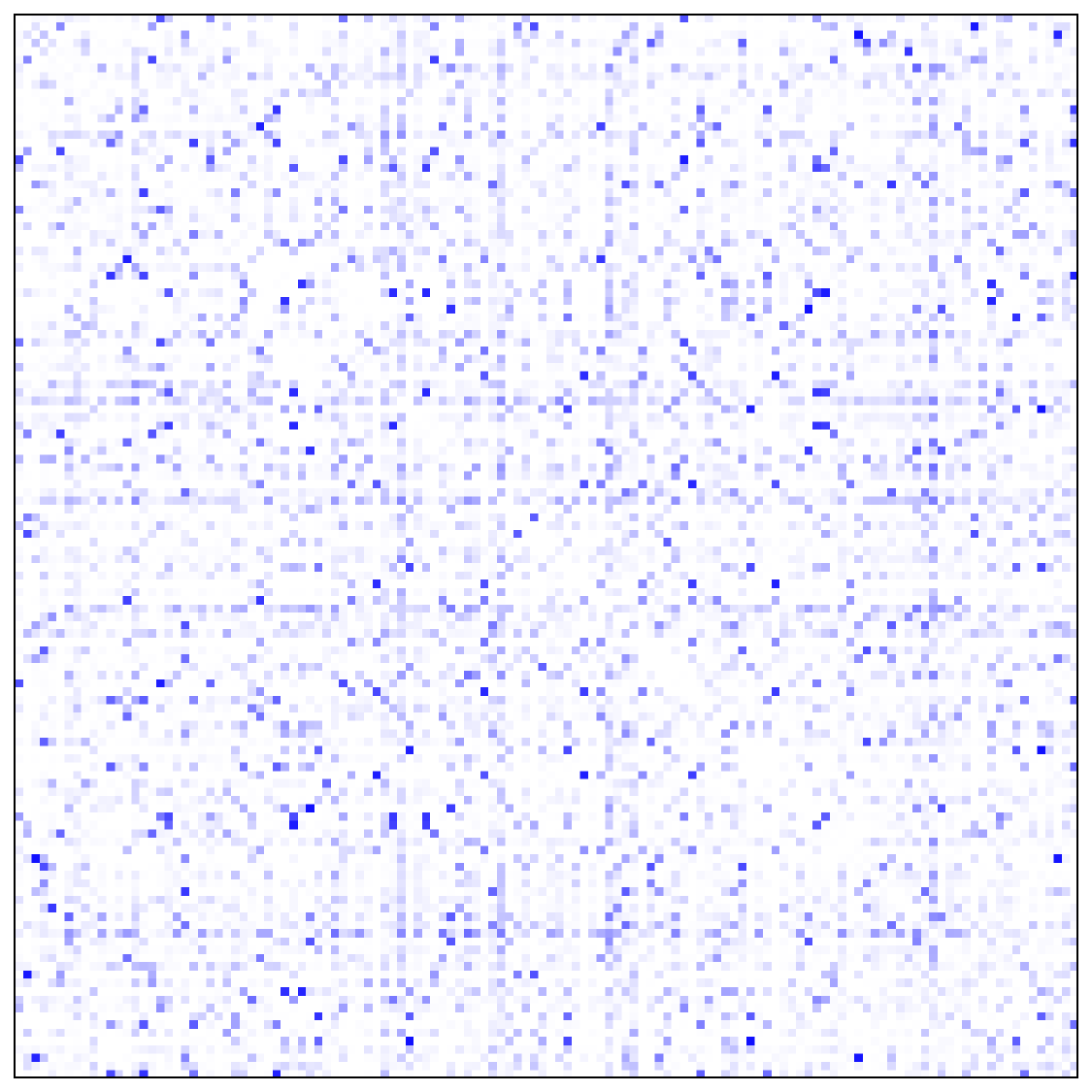}
        \caption{Qwen3-30B-A3B Layer 47}
        \label{fig:coact-qwen}
    \end{subfigure}
    \hfill
    \begin{subfigure}[t]{0.48\linewidth}
        \centering
        \includegraphics[width=\linewidth]{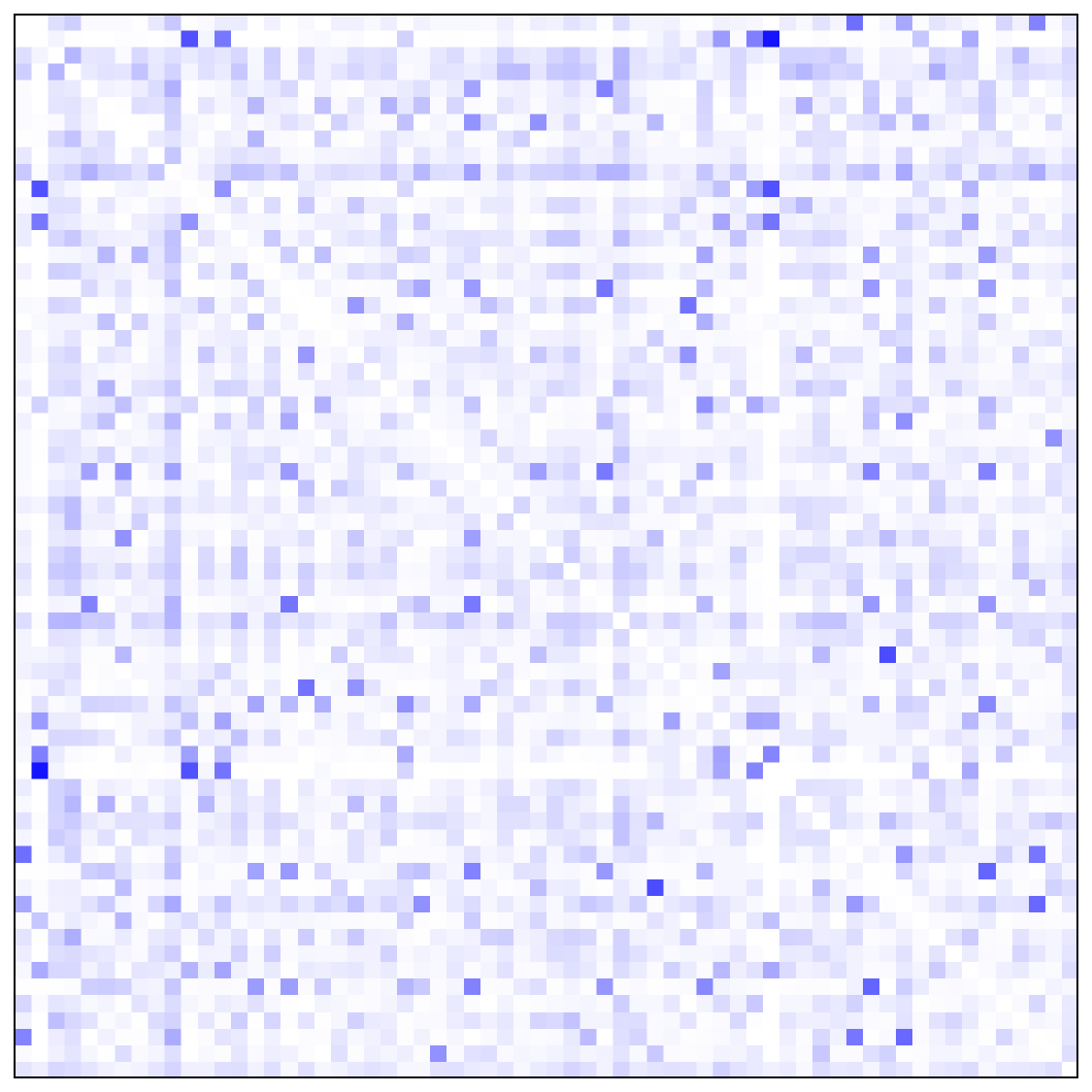}
        \caption{DeepSeek-V2-Lite Layer 24}
        \label{fig:coact-DeepSeek}
    \end{subfigure}
    \caption{Expert co-activation matrix. Each point represents the co-activation frequency of a pair of experts.}
    \label{fig:coactivation_obs}
    
    \vspace{-15pt}
\end{figure}

\begin{figure}[htbp]
  \centering
  \includegraphics[width=\linewidth]{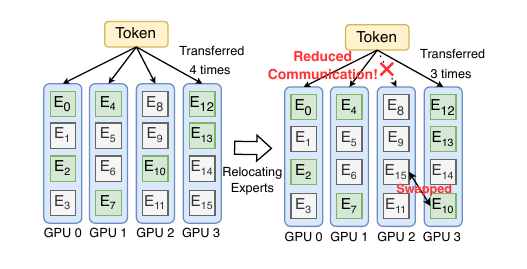}
  
    \vspace{-10pt}
  \caption{Expert placement impacts the communication.}
  \label{fig:reduced_comm}
  
\vspace{-10pt}
  
\end{figure}
In earlier MoE models, the number of experts was relatively small, and each device typically hosted only one expert. In this case, the number of routing destinations of a token was almost fully determined by how many experts were activated, and expert placement had little effect on communication volume. As modern MoE models scale up, however, the number of experts has increased substantially, and each device often needs to host dozens of experts. For example, Qwen3-30B-A3B contains 128 experts per layer, so a single GPU hosts 32 experts when the model is deployed on 4 GPUs.

Under the communication pattern described in Section \ref{sec:bg-comm}, the communication volume depends on the number of distinct devices visited per token, so different expert placements can lead to different communication volumes even under the same routing decisions. As illustrated in Figure~\ref{fig:reduced_comm}, when the activated experts of a token are scattered across multiple devices, the token must be copied and transmitted to each of them, resulting in higher communication overhead. In contrast, if these activated experts are placed on fewer devices, the communication volume is lowered accordingly. This observation is particularly important in modern large-expert-count MoE models. 

Furthermore, we observe that some experts exhibit clear co-activation patterns during inference: if one of them is activated by a token, the others also tend to be activated. If experts that are frequently co-activated can be placed on the same device, the number of routing destinations visited by each token can be reduced.

To validate this observation, we collect pairwise expert co-activation frequencies from inference-time routing traces and visualize the resulting co-activation matrices as Figure \ref{fig:coactivation_obs} shows. Representative examples show that some expert pairs exhibit much stronger co-activation than others. This suggests that expert activations are structured rather than random, and motivates our co-activation-driven placement strategy that groups such experts onto the same device to reduce communication during MoE inference.

\begin{algorithm}
  \caption{Co-activation Driven Expert Placement}
  \label{alg:cadep_adj}
  \KwIn{Number of experts $n$. Number of GPUs $k$. Co-activation probability matrix $P\in\mathbb{R}^{n\times n}$.}
  \KwOut{Expert partition $G$.}
  \SetKw{Return}{Return}
  \SetKwProg{Ft}{Function}{:}{}

  \Ft{} {
    Compute $\deg[i] \leftarrow \sum_{j\ne i} p_{i,j}$ for $i=1..n$ \\
    Initialize $G[g]\leftarrow \emptyset$ for $g=1..k$; \quad $U \leftarrow \{1,2,\dots,n\}$ \\
    Let $\pi$ be experts sorted by $\deg$ in descending order \\
    \For{$g \leftarrow 1$ \KwTo $k$}{
      $s \leftarrow \pi[g]$ \\
      $G[g] \leftarrow G[g]\ \cup\ \{s\}$; \quad $U \leftarrow U \setminus \{s\}$ \\
    }
    $g \leftarrow 1$ \\
    \While{$U \ne \emptyset$}{
      Define $\operatorname{score}(e,S) \leftarrow \sum_{v \in S} p_{e,v}$ \\
      $e^\star \in \arg\max\limits_{e \in U}\ \operatorname{score}(e,G[g])$ \\
      \uIf{$\operatorname{score}(e^\star,G[g]) = 0$}{
        choose $e^\star$ uniformly at random from $U$ \\
      }
      $G[g] \leftarrow G[g]\ \cup\ \{e^\star\}$; \quad $U \leftarrow U \setminus \{e^\star\}$ \\
      $g \leftarrow (g \bmod k) + 1$ \\
    }
    \Return $G[g_1,\dots,g_k]$
  }
\end{algorithm}

\subsection{Co-activation Driven Expert Placement}

\begin{figure}
    \centering
    \includegraphics[width=\linewidth]{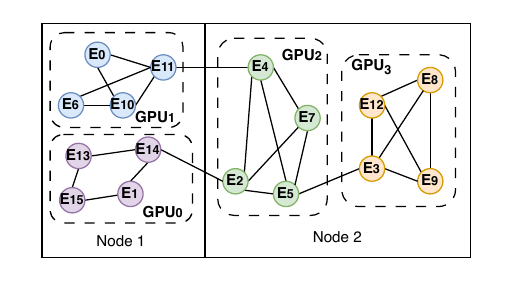}
    \vspace{-20pt}
    \caption{Grouping frequently co-activated experts onto the same devices and nodes using co-activation graph.}
    \label{fig:ap-1}
    \vspace{-20pt}
\end{figure}

To place experts that tend to be co-activated onto the same devices and nodes and thereby reduce communication overhead, we propose \textbf{co-activation driven expert placement} (the \textbf{Placement Approach}). To analyze the co-activation patterns of experts, we first run inference and record the selected experts for each token in each layer. Then for any pair of experts \(i\) and \(j\), we define their co-activation probability \(p_{ij}\) as the ratio between the number of times they are selected together and the total number of times either of them is selected. Based on these statistics, we construct an undirected weighted graph for every layer as Figure \ref{fig:ap-1} shows, where each vertex represents an expert and the edge between experts \(i\) and \(j\) is weighted by \(p_{ij}\).

Let \(\pi(i)\) denote the GPU assigned to expert \(i\). To model communication in cluster deployment, we define the communication objective as
\begin{equation}
\mathcal{C}(\pi)=\sum_{i<j} p_{ij}\, d(\pi(i),\pi(j))
\end{equation}
where \(d(\pi(i),\pi(j))\) denotes the communication cost between the GPUs hosting experts \(i\) and \(j\). Specifically,
\begin{equation}
d(\pi(i),\pi(j))=
\begin{cases}
0, & \pi(i)=\pi(j),\\
1, & \pi(i)\neq\pi(j) \text{ and } \mathrm{node}(\pi(i))=\mathrm{node}(\pi(j)),\\
c, & \mathrm{node}(\pi(i))\neq\mathrm{node}(\pi(j))
\end{cases}
\end{equation}
, where $c$ is the relative communication cost between nodes. Accordingly, the placement problem is to find $\pi$ that minimizes $\mathcal{C}(\pi)$,
subject to the constraint that each GPU hosts the same number of experts. Under this formulation, communication minimization becomes a capacity-constrained weighted graph partitioning problem, which is combinatorial and expensive to solve exactly at inference-system scale. We therefore adopt a two-stage greedy algorithm to approximately optimize this objective.

In the first stage, we construct GPU-level expert groups. As Algorithm~\ref{alg:cadep_adj} shows, we first compute the weighted degree $deg[i]$ of each expert and select the \(k\) experts with the largest weighted degrees as seeds, assigning one seed to each GPU group $G[g]$. Starting from these seeds, the algorithm iteratively grows the \(k\) groups in a round-robin manner. In each step, for the current GPU, it selects the unassigned expert with the largest cumulative co-activation weight to the experts already assigned to that GPU. This process continues until all experts are assigned. 

In the second stage, we further assign GPU groups to nodes. After constructing GPU-level expert groups, we treat each GPU group as a coarse-grained unit, compute the co-activation strength between groups, and apply the same grouping principle again to assign GPU groups to nodes. In this way, the first stage reduces communication across GPUs, while the second stage further reduces communication across nodes, which reduces search complexity compared with directly optimizing the weighted graph partitioning problem. These stages will be performed on every MoE layer to generate an expert placement for the entire model.

This hierarchical placement reduces the average number of devices each token needs to interact with, thereby reducing the communication overhead of expert parallelism and alleviating expensive cross-device and cross-node transfers. Meanwhile, the algorithm ensures that each device hosts the same number of experts, keeping expert weight memory and KV-cache capacity balanced across devices. This balance prevents any single device from becoming a memory bottleneck, which is especially important for long-sequence, large-model inference.

\section{Comm.-comp. Trade-off Adjustment}

\label{sec:ap-2}

\begin{figure}
    \centering
    \includegraphics[width=0.85\linewidth]{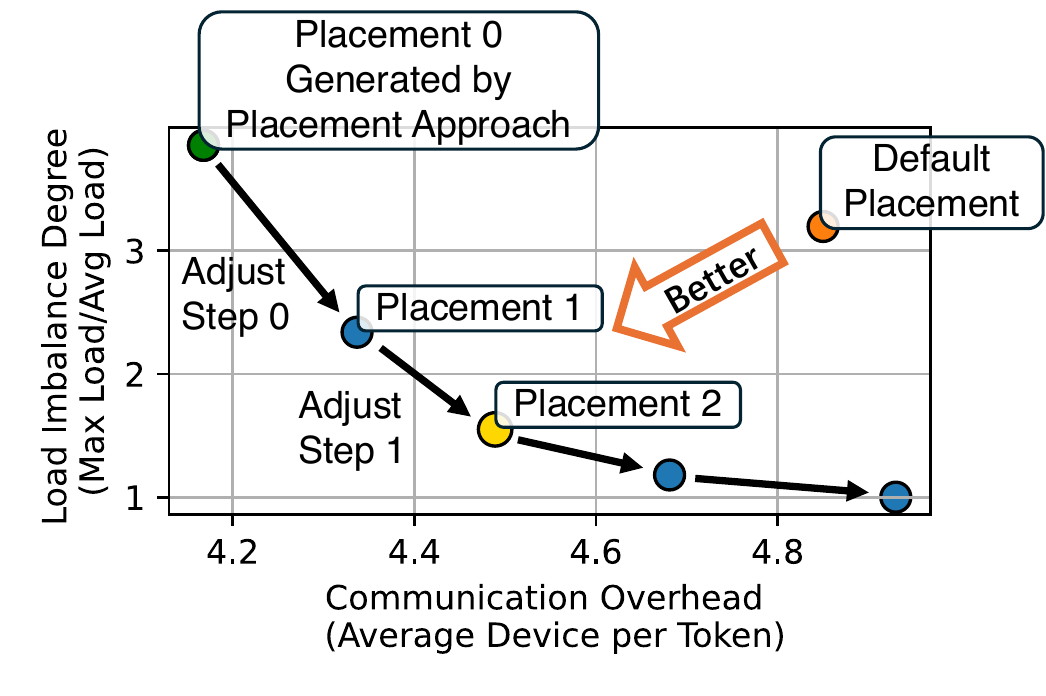}
    
    \vspace{-10pt}
    \caption{Comm. \& Comp. Trade-off Adjustment}
    
    \label{fig:ap2}
    
    \vspace{-20pt}
\end{figure}

\subsection{Trade-off between Communication and Load Balance}

The Placement Approach in Section \ref{sec:ap-1} is communication-oriented: it reduces communication by grouping frequently co-activated experts onto the same device. However, minimizing communication alone is not sufficient for optimizing end-to-end performance. Since hot experts may also be co-activated experts, placing them on the same device can concentrate hot experts on few GPUs and lead to substantial load imbalance, which is not necessarily the best placement in terms of overall system performance.

This reveals an inherent trade-off between communication and load balance. A placement with lower communication tends to colocate more strongly co-activated experts, but such colocation may also make device loads more uneven. Conversely, a more balanced placement often requires separating some co-activated experts, which increases communication. Therefore, optimizing only communication or only load balance is insufficient.

Moreover, the preferred trade-off depends on the characteristics of the underlying hardware, since the relative communication and computation capabilities vary across different machines. This observation motivates us to formulate expert placement as a communication–load multi-objective optimization problem.

\subsection{Modeling and Optimization}
\begin{figure*}
    \centering
    \includegraphics[width=\linewidth]{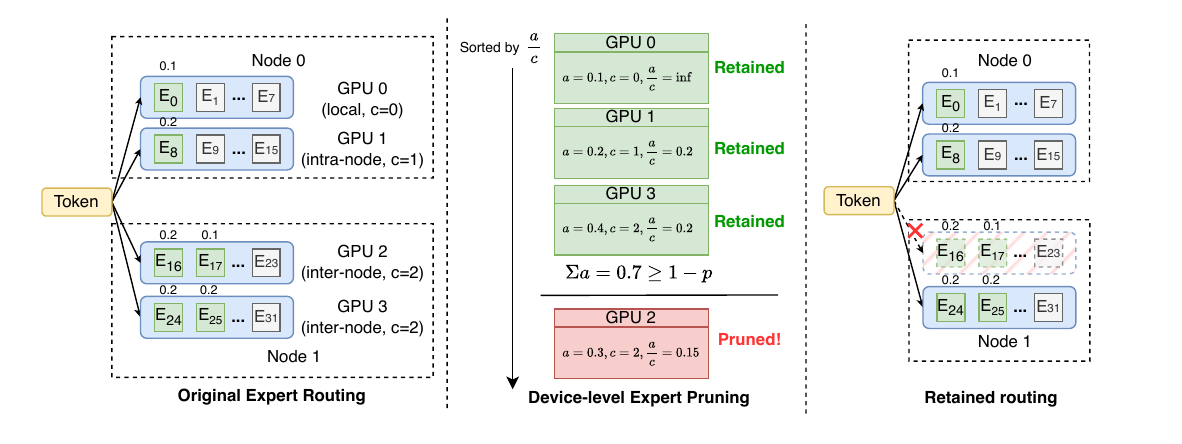}
    \vspace{-25pt}
    \caption{Communication-aware pruning with a 0.3 threshold}
    \vspace{-10pt}
    \label{fig:pruning}
\end{figure*}

Based on the above observation, we define the load imbalance objective as
\begin{equation}
\mathcal{B}(\pi)=\frac{\max_{g} L_g}{\frac{1}{k}\sum_{g} L_g}
\end{equation}
, where \(L_g\) denotes the number of tokens processed by GPU \(g\), $k$ denotes the number of GPUs. A smaller \(\mathcal{B}(\pi)\) indicates a more balanced placement. Combined with the communication objective \(\mathcal{C}(\pi)\) defined in Section~\ref{sec:ap-1}, we consider the following communication-load objective:
\begin{equation}
\min_{\pi}\ \lambda \mathcal{C}(\pi) + (1-\lambda)\mathcal{B}(\pi)
\end{equation}
, where \(\lambda \in [0,1]\) controls the trade-off between communication cost and load balance. A larger \(\lambda\) places more emphasis on minimizing communication, while a smaller \(\lambda\) places more emphasis on improving load balance.

Directly solving this objective remains difficult, so we propose a \textbf{Comm.-comp. Trade-off Adjustment Approach} (the \textbf{Adjustment Approach}). We start from the communication-oriented placement produced by the Placement Approach and iteratively refine it through expert swaps. In each adjustment step, we swap a pair of experts between high-load and low-load devices to reduce load imbalance while preserving as much co-activation locality as possible. Starting from the communication-oriented placement, this process generates a placement spectrum with gradually increasing communication cost and improving load balance.

To illustrate the trade-off among different expert placements, we construct the visualization shown in Figure~\ref{fig:ap2}. Each expert placement corresponds to a single point in this space. The x-axis indicates the average number of devices each token is routed to under a given placement, reflecting its communication volume. The y-axis measures load imbalance across devices according to \(\mathcal{B}(\pi)\). Starting from the placement produced by the Placement Approach---the green point in the upper-left corner with minimal communication but high load imbalance---the Adjustment Approach iteratively generates new placements that move toward lower imbalance at the cost of higher communication. The resulting placement spectrum can be interpreted as approximate solutions to the communication-load objective under different trade-off weights \(\lambda\).

The optimal choice depends on the characteristics of the underlying hardware. Comm.-rich machines have abundant interconnect bandwidth and therefore prefer more load-balanced placements, even if they incur higher communication. Comm.-constrained machines suffer from limited bandwidth, where aggressively pursuing load balance leads to excessive communication overhead that outweighs computational gains. Balanced machines exhibit comparable compute and communication capabilities, and therefore prefer placements in the middle of the spectrum. Using lightweight performance profiling, CAP automatically selects the most suitable placement for each hardware type.

Furthermore, as shown in Figure~\ref{fig:ap2}, our experimental results show that the placement spectrum produced by the Adjustment Approach forms a Pareto frontier over the widely-used default sequential placement, which assigns experts to devices strictly by numerical expert ID order. That means, multiple placements in our spectrum achieve both lower communication cost and better load balance than the default placement. 

\section{Communication-Aware Expert Pruning}

Dynamic expert pruning is a common technique for reducing the computational cost of MoE inference. During inference, it further reduces the number of experts actually involved in computation for each token according to token-dependent routing scores. Traditional dynamic pruning methods do not explicitly consider how different pruning decisions affect communication overhead. In particular, they do not distinguish whether pruning removes inexpensive intra-node communication or more expensive inter-node communication. To address this, we propose \textbf{Communication-Aware Expert Pruning} (the \textbf{Pruning Approach}), which explicitly incorporates communication cost into the pruning decision.

Rather than pruning experts independently, we perform pruning at the device level. This design is motivated by the fact that pruning an entire device can eliminate one routing destination at once, yielding a much larger communication reduction than fragmented expert-level pruning.

To model this process, we associate each device $d$ with two quantities. The first is its accuracy contribution $a_d$, which is the sum of routing scores of all candidate experts on that device. We use this quantity as a proxy for accuracy contribution, because the final MoE output is computed as a weighted aggregation of expert outputs as Eq. \ref{eq:moe} shows, and the routing scores directly determine how much each retained expert contributes to the token output. This is also consistent with prior expert pruning methods that use routing scores to estimate expert importance. The second is its communication cost $c_d$, which is determined by the topology between the current device of the token and the target device, including whether the communication stays on the same device, crosses GPUs within a node, or crosses nodes. Specifically, devices corresponding to intra-node communication are assigned a cost of $1$, while devices corresponding to inter-node communication are assigned a larger predefined constant $c$. Let $x_d \in \{0,1\}$ indicate whether device d is retained. We then formulate pruning as selecting a set of devices with minimum communication cost while preserving sufficient accuracy contribution:
\begin{equation}
    \min_{x_d \in \{0,1\}} \sum_d c_d x_d
    \qquad
    \text{s.t.}\qquad
    \sum_d a_d x_d \ge 1 - p
    \label{eq:pruning}
\end{equation}
, where $p$ is a user-specified pruning threshold that controls the accuracy–performance trade-off, while CAP optimizes inference under the given threshold. In practice, $c$ can be chosen according to the ratio between inter-node and intra-node link bandwidths.

Although Eq. \ref{eq:pruning} can be viewed as a cost-constrained discrete selection problem, we adopt a lightweight greedy solver because it is efficient for online pruning and sufficient for choosing experts to prune. As Figure \ref{fig:pruning} shows, we first compute the device-level accuracy contribution $a_d$. We then compute the ratio $\frac{a_d}{c_d}$ for each device and retain devices in descending order of this ratio until the target accuracy $1 - p$ is satisfied. We implement GPU-friendly vectorized kernels, making the overhead of pruning during inference negligible. This device-level pruning strategy removes routing destinations, which yields larger communication reduction than fragmented expert pruning. As a result, our pruning approach can achieve higher end-to-end speedup under the same pruning threshold, or equivalently, preserve better model accuracy under the same target speedup.
\begin{figure*}[ht]
    \centering
    \includegraphics[width=\linewidth]{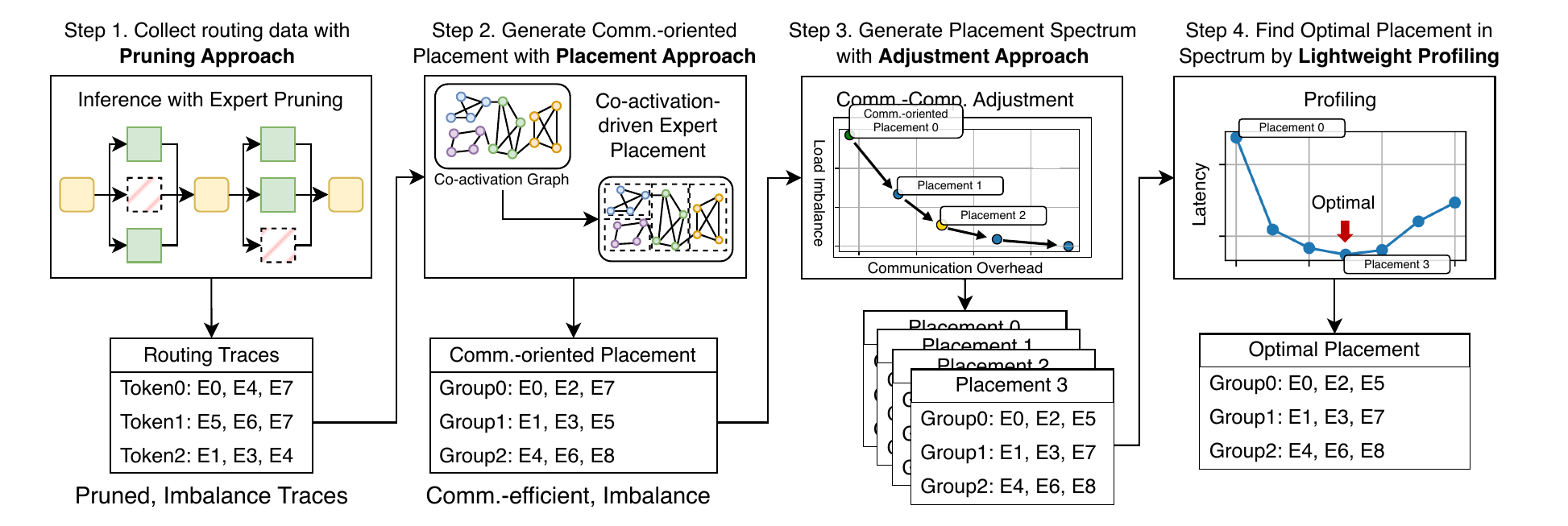}
    
    \caption{The 4-step Pipeline of CAP}
    
    \vspace{-20pt}
    \label{fig:overall}
\end{figure*}

\section{Overall Pipeline}
\label{sec:overall-pipeline}

The three components of CAP serve different purposes. The Placement Approach reduces communication by grouping experts that tend to be co-activated onto the same device. The Adjustment Approach further explores different trade-offs between communication cost and load balance. The Pruning Approach reduces computation and communication simultaneously by removing routing destinations with low contribution.

\begin{figure}[t]
    \centering
    \begin{minipage}[t]{0.45\linewidth}
        \centering
        \includegraphics[width=\linewidth]{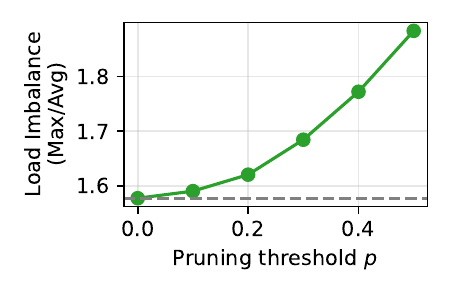}
        \captionof{figure}{Load balance under different pruning thresholds.}
        \label{fig:p-load}
    \end{minipage}
    \hfill
    \begin{minipage}[t]{0.45\linewidth}
        \centering
        \includegraphics[width=\linewidth]{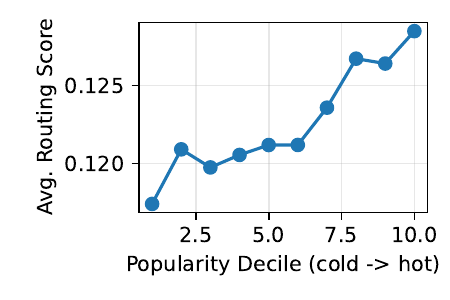}
        \captionof{figure}{Hot experts tend to get higher routing scores.}
        \label{fig:pop-scores}
    \end{minipage}
    
    \vspace{-20pt}
\end{figure}

These three components should be applied in a specific order. First, pruning changes the expert co-activation pattern by removing part of the routing destinations. Therefore, the Placement Approach should be performed on the pruned routing data rather than on the original one. 

Second, our preliminary results show that pruning also tends to increase load imbalance as Figure \ref{fig:p-load} shows. We statistically analyze the routing score distributions of experts with different popularity, and observe that cold experts usually have lower routing scores when activated and are therefore more likely to be pruned as Figure \ref{fig:pop-scores} shows. This makes the remaining routing destinations more concentrated on hot experts and leads to a more imbalanced load distribution.

As a result, after applying the Placement Approach on the pruned routing data, CAP obtains an initial grouping with lower communication but higher load imbalance. This is exactly the regime targeted by the Adjustment Approach, which starts from such a communication-efficient but imbalanced grouping and generates a sequence of placements with different communication-load balance trade-offs.

Accordingly, CAP is executed in four steps as Figure \ref{fig:overall} shows. We first run the system with communication-aware expert pruning enabled, and collect token-level routing traces. Then, in Step 2, based on the pruned routing data, we perform co-activation-driven expert placement to obtain a communication-oriented initial grouping. In Step 3, we apply the comm.-comp. trade-off adjustment method to refine this initial grouping and generate a placement spectrum with different communication-load balance trade-offs. In the final step, lightweight profiling is then used to select the most suitable placement for the target hardware platform.

\section{Evaluation}
\label{sec:evaluation}
\subsection{Experiment Setup}
\paragraph{Testbed} We evaluate CAP on three machines, each representing one of the compute–communication characteristics discussed earlier in Section \ref{sec:ap-2}. We conduct experiments on all three nodes to demonstrate CAP's strong hardware adaptivity.
{\setlength{\leftmargini}{10pt}
\begin{itemize}
    \item Node A is equipped with eight RTX 3090 GPUs. It lacks NVLink and GPU-Direct P2P support. All communication must be routed through the CPU, making it a comm.-constrained platform.
    \item Node B contains eight A100 GPUs. Although it lacks NVLink, it can adopt GPU-Direct P2P over PCIe. We refer to it as a comm.-comp. balanced node.
    \item Node C features eight H100 GPUs connected via NVLink, corresponding to a comm.-rich platform.
\end{itemize}
}

To further evaluate the scalability of CAP on multi-node clusters and its ability to reduce inter-node communication, we conduct experiments on two cluster configurations.
\begin{itemize}
    \item Cluster A consists of 2 nodes with 8 H100 GPUs per node with NVLinks, where the two nodes are connected by 8$\times$400 Gbps InfiniBand. We report the end-to-end performance on Cluster A.
    \item Cluster B also consists of 2 nodes with 8 H100 GPUs per node with NVLinks, but the inter-node connection bandwidth is reduced to only 200 Gbps. We use this cluster to analyze the impact of inter-node communication cost under different bandwidth conditions in Section \ref{sec:ev-cluster}.
\end{itemize}

\paragraph{Model Setup} We select two MoE models from different model families: Qwen3-30B-A3B and DeepSeek-V2-Lite. They contain 128 and 64 experts respectively, representing the latest generation of MoE models with large expert counts.

\paragraph{Baselines} 
{\setlength{\leftmargini}{10pt}

We compare CAP against two baselines that cover the main deployment and optimization choices in existing MoE inference systems.

\textbf{Default} assigns experts by sequential expert ID order, which corresponds to the standard deployment used in mainstream inference frameworks such as \textbf{vLLM}\cite{kwon2023efficient} and \textbf{SGLang}\cite{zheng2024sglangefficientexecutionstructured}. 

\textbf{DeepSeek EPLB} is a stronger algorithmic baseline that explicitly optimizes expert placement for load balance. It is representative of load-balance-oriented placement, which is the main optimization target of prior work.

Together, these two baselines span the two most relevant comparison points for this work: real-world default deployment and state-of-the-art load-balance-oriented placement.
}
\begin{figure*}
    \centering
    \includegraphics[width=0.98\linewidth]{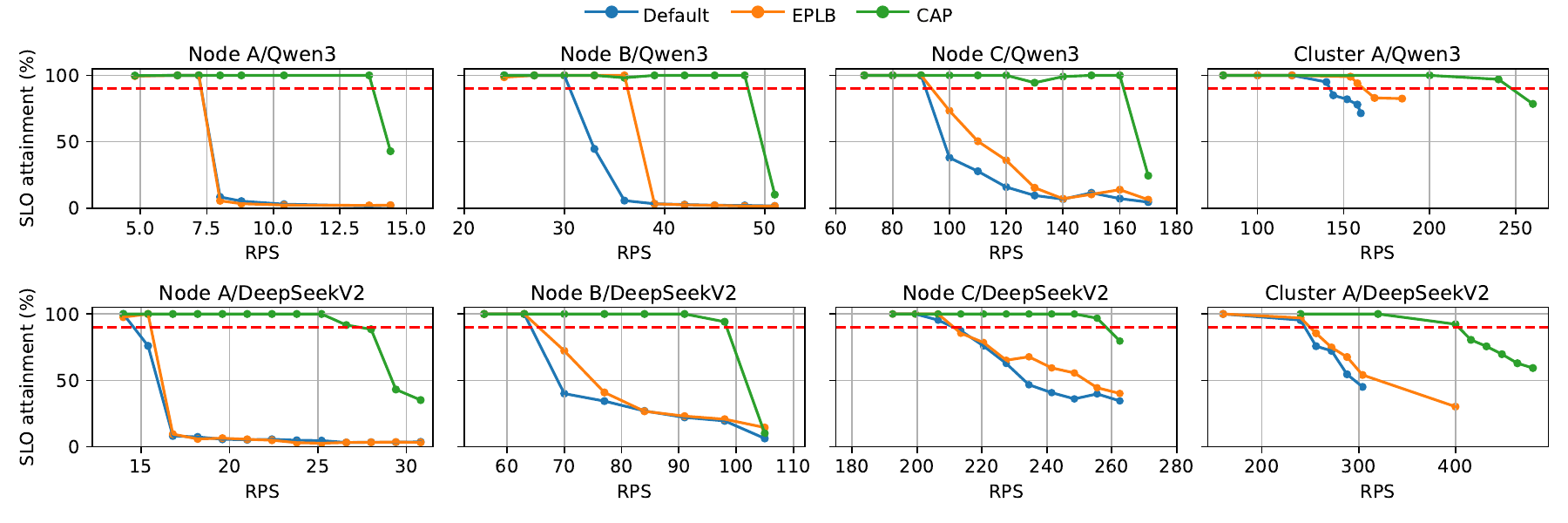}
    \vspace{-10pt}
    \caption{End-to-end Performance}
    \label{fig:e2e}
    
    \vspace{-15pt}
\end{figure*}
\paragraph{Dataset and Metrics} For performance evaluation, we use the LMSYS-Chat and Arxiv Abstracts datasets in our evaluations. For each node, we manually select the optimal batch size and generate requests at different input rates by sampling from the dataset. We adopt 6 s TTFT and 500 ms TBT as the service-level objectives (SLOs). During end-to-end testing, we classify a configuration as violating the SLO if fewer than 90\% of the requests meet these thresholds. For accuracy evaluation, we use HumanEval, MMLU and GSM8K to evaluate CAP accuracy under different expert pruning thresholds.

\subsection{End-to-end Performance}

We compare CAP with the two baselines across different testbeds with different models by measuring their end-to-end throughput.
We replay requests under increasing input rates and report the maximum throughput that still satisfies the SLOs.
CAP treats pruning threshold $p$ as a user-controlled accuracy–performance knob rather than a tuned internal hyperparameter. In this evaluation, we report results under several representative values of $p$ to illustrate the trade-off between accuracy and speedup. We set $p$ to 0.3 for Qwen3 and 0.1 for DeepSeek-V2-Lite, under which the perplexity increase of both models remains about 10\%, a level that prior work ~\cite{huang2025discovering} suggests still preserves strong accuracy. Detailed accuracy evaluation is in Section \ref{sec:eva-pruning}.

Across all testbeds, CAP consistently outperforms the baselines as Figure \ref{fig:e2e} shows, with the largest gains observed on the comm.-constrained node, where it achieves up to 1.86$\times$ throughput improvement. This advantage arises because CAP is the only method that explicitly accounts for communication volume, which dominates total latency on comm.-constrained systems.
On the other testbeds, CAP also identifies the best-performing configuration by combining expert pruning with its comm.–comp. trade-off analysis, achieving the best throughput on all tested nodes.

EPLB improves over the Default placement on the balanced and comm.-rich machines but performs similarly to, or even worse than Default on the comm.-constrained machine.
Although EPLB achieves better load balance, it incurs significantly higher communication volume, thereby causing its communication overhead to exceed the computational gains on low-bandwidth hardware.
This highlights that, in MoE inference, load balance alone is insufficient; communication volume is likewise a critical determinant of end-to-end performance.

\subsection{Analysis on the Placement Spectrum}

\begin{figure}
    \centering
    \includegraphics[width=0.98\linewidth]{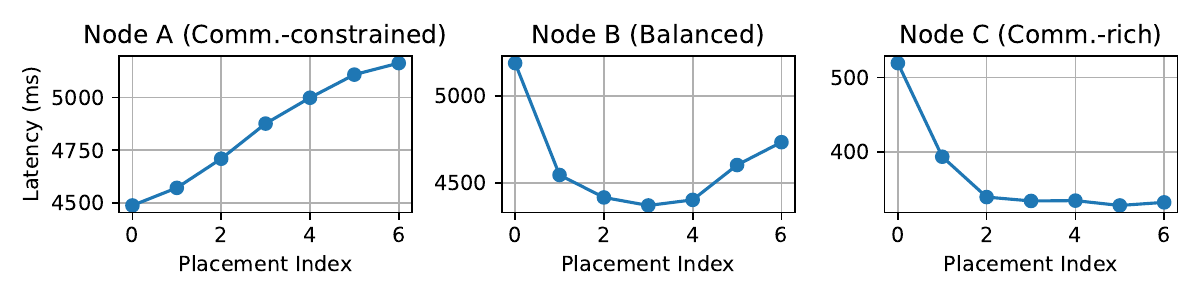}
    \caption{Placement Spectrum Performance on 3 Nodes}
    \label{fig:ev-adj}
    
\end{figure}

\begin{figure}[htbp]
\centering
\begin{minipage}[t]{0.48\linewidth}
    \centering
    \includegraphics[width=\textwidth]{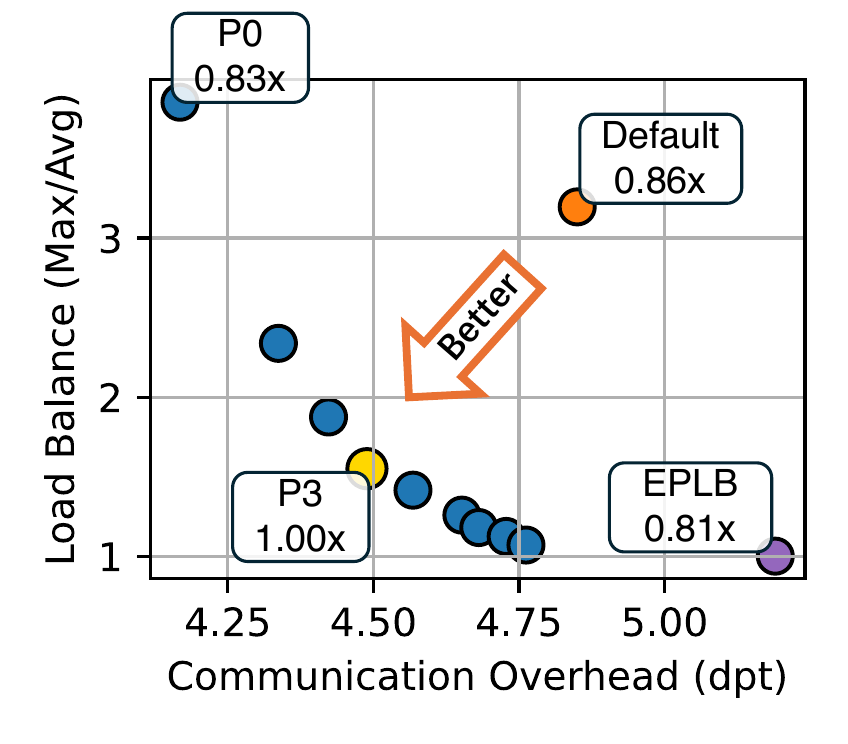}
    \caption{Performance and Communication-Computation Characteristics of Placement Spectrum}
    \label{fig:ev-theory}
\end{minipage}
\hfill
\begin{minipage}[t]{0.48\linewidth}
    \centering
    \includegraphics[width=\textwidth]{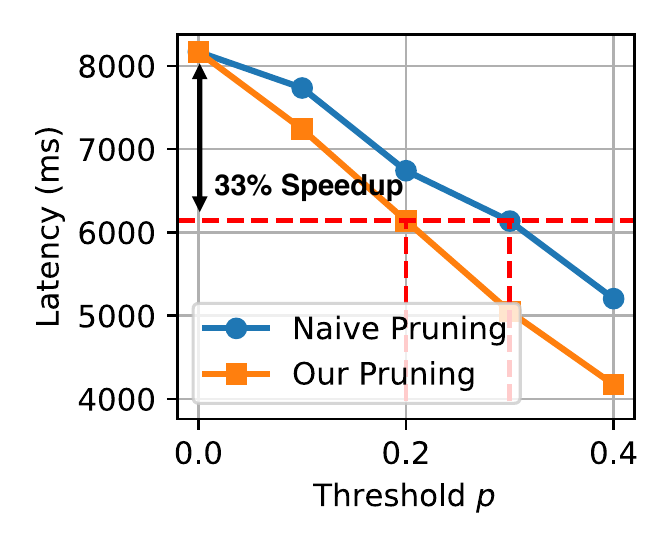}
    
    \caption{Performance of CAP and Naive Pruning under Different Thresholds}
    \label{fig:ev3}
\end{minipage}
\vspace{-20pt}
\end{figure}

We conduct experiments to analyze the performance characteristics of the placement spectrum generated by the Adjustment Approach on different hardware platforms.
As shown in Figure \ref{fig:ev-adj}, we evaluate the latency of every placement in the spectrum generated for Qwen3-30B-A3B on each node with the same pruning threshold of $p=0.3$. 
Placements with lower indices exhibit lower communication volume but higher load imbalance (corresponding to the upper-left region in Figures \ref{fig:ap2} and \ref{fig:ev-theory}).
On the comm.-constrained node, communication dominates the overall latency, making the communication-minimal placement the optimal choice. 
On the balanced node, a mid-spectrum placement (e.g., Placement 3), which provides a more favorable communication–load trade-off, achieves the best performance. 
On the comm.-rich node, where communication is extremely fast, the most load-balanced placements deliver the best performance. These results show that optimal MoE inference performance requires hardware-aware selection from the communication–load trade-off spectrum rather than optimizing either dimension alone.

We further examine the performance differences among placements on Node B (the balanced machine). On this device, placement 3 achieves the best performance. 
It delivers a 20.4\% improvement over the unadjusted placement produced by the Placement Approach, a 23.4\% improvement over the load balanced but high-communication placement produced by EPLB, and a 16\% improvement over the Default placement, demonstrating that the Adjustment Approach is effective on its own.

\subsection{Effectiveness of the Pruning Approach}
\label{sec:eva-pruning}

\begin{table}[t]
\centering
\caption{Accuracy under different pruning thresholds. Lower is better for PPL, while higher is better for others. }
\label{tab:pruning-accuracy}
\begin{tabular}{lccc}
\toprule
Metric & \(p=0\) & \(p=0.2\) & \(p=0.3\) \\
\midrule
PPL $\downarrow$       & 17.36  & 18.17
{\color{BrickRed}(+0.81)} & 19.12 {\color{BrickRed}(+1.76)} \\
GSM8K $\uparrow$       & 0.8863 & 0.8590 {\color{BrickRed}(-2.73 pp)} & 0.8514 {\color{BrickRed}(-3.49 pp)} \\
HumanEval $\uparrow$   & 0.8902 & 0.8841 {\color{BrickRed}(-0.61 pp)} & 0.8598 {\color{BrickRed}(-3.04 pp)} \\
MMLU $\uparrow$        & 0.7154 & 0.7161 {\color{ForestGreen}(+0.07 pp)} & 0.6874 {\color{BrickRed}(-2.80 pp)} \\
\bottomrule
\end{tabular}
\vspace{-15pt}
\end{table}

To evaluate the effectiveness of communication-aware expert pruning, we compare CAP with naive dynamic pruning under the same pruning threshold on Node A. The threshold has the same semantics in both methods: for each token, they prune experts whose cumulative routing scores are less than a predefined threshold \(p\). Therefore, under the same \(p\), the two methods apply a similar pruning strength. The key difference is that CAP additionally considers communication cost during pruning, while naive pruning only reduces computation.

As a result, as Figure \ref{fig:ev3} shows, under the same pruning threshold, CAP achieves higher latency speedup than naive pruning because it reduces both computation and communication. In our experiments, CAP with \(p=0.2\) achieves the same 33\% latency reduction as naive pruning with \(p=0.3\). Since a smaller threshold retains more routing destinations and prunes fewer experts, CAP can preserve higher model accuracy under the same target speedup.

To quantify this effect, we further evaluate the accuracy of models under different pruning thresholds. Table~\ref{tab:pruning-accuracy} reports the results at \(p=0\), \(0.2\), and \(0.3\) for MMLU, GSM8K, HumanEval and perplexity. Across all benchmarks, increasing the pruning threshold from \(p=0.2\) to \(p=0.3\) consistently leads to larger accuracy degradation. These results indicate that the additional pruning required by a larger threshold causes accuracy loss across knowledge, reasoning, and code generation tasks.

Overall, the main advantage of communication-aware pruning is that it not only accelerates inference more effectively but also preserves better accuracy under the same performance requirements.

\begin{figure*}[ht]
    \centering
    \includegraphics[width=\linewidth]{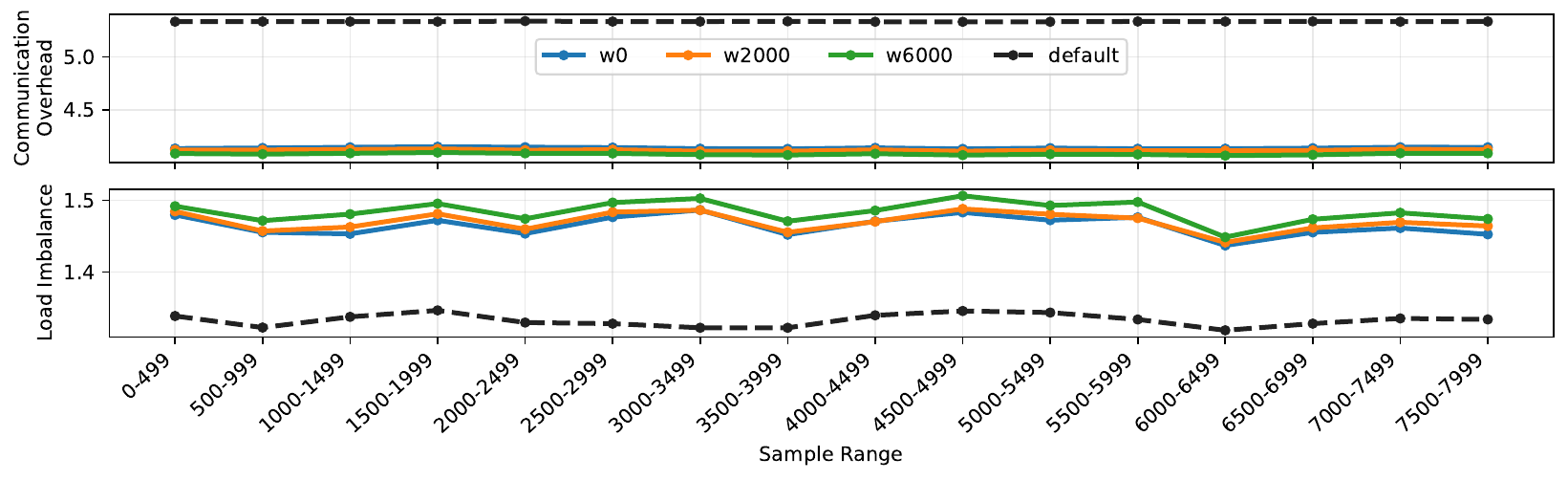}
    \caption{In-workload stability of expert grouping on LMSYS Chat. Groupings generated from three different windows (w0, w2000, and w6000) are evaluated on all 16 windows of all 8000 requests.}
    \label{fig:in_domain_stability}
    \vspace{-10pt}
\end{figure*}

\subsection{Analysis on Inter-node Expert Pruning in Clusters}
\label{sec:ev-cluster}

In multi-node pruning, the relative cost assigned to inter-node communication affects which routing destinations are preferred for removal. To study this effect, we compare two settings in the pruning objective: \(c=1\), where intra-node and inter-node communication are treated equally, and \(c=5\), where inter-node communication is assigned a much higher cost.

As shown in Table~\ref{tab:inter-node-cost}, increasing the inter-node communication cost indeed reduces cross-node communication. On both clusters, the \(c=5\) setting causes each token to visit fewer nodes than \(c=1\), indicating that pruning becomes more inclined to remove routing destinations that span more nodes.

However, the performance impact of this reduction depends on the actual inter-node bandwidth of the cluster. On Cluster A, where the two nodes are connected by \(8\times400\,\mathrm{Gbps}\) InfiniBand links, reducing the number of nodes visited by each token does not lead to a noticeable latency improvement. In contrast, on Cluster B, where the inter-node bandwidth is only \(200\,\mathrm{Gbps}\), the same change leads to a clear latency reduction. This result shows that topology-aware pruning can effectively reduce inter-node communication, and its benefit becomes more significant when inter-node bandwidth is limited. In practice, $c$ can be set to reflect the relative communication cost between intra-node and inter-node links.

\begin{table}[t]
\centering
\caption{Impact of inter-node communication cost in multi-node pruning}
\label{tab:inter-node-cost}
\begin{tabular}{lccc}
\toprule
Cluster & Config & Latency (ms) & Node per Token \\
\midrule
\multirow{2}{*}{A (8$\times$400\,Gbps)}
& $c=1$ & 1434.68 & 1.8875 \\
& $c=5$ & 1447.89 {\color{BrickRed}(+0.92\%)} & 1.8337 {\color{ForestGreen}(-2.85\%)} \\
\midrule
\multirow{2}{*}{B (200\,Gbps)}
& $c=1$ & 2285.37 & 1.8875 \\
& $c=5$ & 1980.26 {\color{ForestGreen}(-13.35\%)} & 1.8337 {\color{ForestGreen}(-2.85\%)} \\
\bottomrule
\end{tabular}
\vspace{-20pt}
\end{table}

\subsection{Stability and Transferability of Expert Grouping}
\label{sec:grouping-transfer}

To study the stability of expert grouping, we conduct both in-workload and cross-workload experiments.

We first evaluate in-workload stability on the LMSYS Chat dataset. We take the first 8000 requests and divide them into 16 windows of 500 requests each according to their original order. We then use three different windows of requests, namely requests 0--499 (w0), 2000--2499 (w2000), and 6000--6499 (w6000), to run inference and generate communication-oriented groupings using the Placement Approach. Then we evaluate each of these groupings on all 16 windows. The results are shown in Figure~\ref{fig:in_domain_stability}. 

Both communication cost and load imbalance remain stable across all windows for all three groupings. Specifically, the maximum variation is below 2.9\% for communication cost and below 5.8\% for load imbalance. Communication cost also remains consistently lower than that of the default placement. These results show that the grouping produced by the Placement Approach is stable within the same workload.  

We next evaluate cross-workload transferability using three datasets: HumanEval, LMSYS Chat, and Arxiv Abstracts. For each dataset, we generate a communication-oriented grouping and evaluate all groupings on all three datasets. Table~\ref{tab:cross-domain-dpt} and Table~\ref{tab:cross-domain-m2a} show that both communication cost and load imbalance remain relatively stable across source-target combinations, indicating strong cross-domain transferability.

Overall, these results show that expert grouping exhibits strong stability within the same workload and strong transferability across different workloads. Therefore, the need to frequently regroup experts is limited in practice.

\subsection{Optimization Overhead}
CAP requires an offline optimization phase before deployment to generate and adjust expert groupings. This process consists of two stages. The first stage generates the candidate groupings, including collecting routing statistics, constructing the initial placement, and performing adjustment. This stage is hardware-independent and only needs to be executed once for a given model. To keep this process efficient, we implement both expert grouping and adjustment with GPU-efficient vectorized code. In our experiments, we find that the routing statistics and the resulting groupings become stable after collecting about 8,000 tokens. On an 8-A100 machine with Qwen3, collecting these tokens takes 36.72 seconds, generating the initial placement takes about 80 seconds, and producing 7 adjusted placements takes 9.1 seconds.

The second stage profiles all candidate placements on a target machine and selects the best one. Using the same 8,000-token workload, evaluating 7 placements takes about 240 seconds. Overall, the full optimization overhead is \textbf{no more than 6 minutes}, which we consider acceptable in practice. Moreover, grouping generation is only required once per model, and placement profiling is only required once per machine, so the amortized cost is even lower. As discussed in Section~\ref{sec:grouping-transfer}, the need for frequent regrouping is also limited in practice.

\begin{table}[t]
\centering
\caption{Cross-workload \textbf{communication cost} measured by Device per Token. Rows denote target workloads and columns denote the workload used for profiling the grouping.}
\label{tab:cross-domain-dpt}
\begin{tabular}{lccc|c}
\toprule
Target $\backslash$ Source & HumanEval & LMSYS Chat & Arxiv & Default \\
\midrule
HumanEval & \textbf{4.1753} & 4.2481 & 4.4817 & 5.3497 \\
LMSYS Chat & 4.4444 & \textbf{4.1405} & 4.4223 & 5.3315 \\
Arxiv & 4.4930 & \textbf{4.2832} & 4.3443 & 5.3408 \\
\bottomrule
\end{tabular}
\vspace{-10pt}
\end{table}

\begin{table}[t]
\centering
\caption{Cross-workload \textbf{load imbalance} measured by Max Load/Avg. Load. }
\label{tab:cross-domain-m2a}
\begin{tabular}{lccc|c}
\toprule
Target $\backslash$ Source & HumanEval & LMSYS Chat & Arxiv & Default \\
\midrule
HumanEval & 1.6772 & 1.8069 & \textbf{1.6629} & 1.4372 \\
LMSYS Chat & 1.5620 & \textbf{1.4675} & 1.5008 & 1.3303 \\
Arxiv & 1.8140 & 1.9201 & \textbf{1.6741} & 1.5972 \\
\bottomrule
\end{tabular}
\vspace{-10pt}
\end{table}

\section{Related Work}
Existing efforts to improve MoE inference mainly fall into two categories: load-balancing methods that optimize expert placement or routing, and expert pruning methods that reduce computation by removing low-contribution experts.
\paragraph{MoE Inference Load Balancing} Load imbalance is a major issue in MoE inference\cite{han2025gracemoegroupingreplicationlocalityaware,huang2024toward,yu2024moesysdistributedefficientmixtureofexperts,li2024acceleratingdistributedmoetraining,rajbhandari2022deepspeedmoeadvancingmixtureofexpertsinference,Nie_2023,moetuner}. At the algorithmic level, GShard \cite{lepikhin2020gshard} introduces an auxiliary loss to encourage more even token routing, and Switch Transformer\cite{fedus2022switchtransformersscalingtrillion} constrains expert capacity to limit overload, though both approaches may reduce routing flexibility and weaken model expressiveness. System-level solutions have also been explored: FasterMoE \cite{he2022fastermoe} uses shadow experts to clone hot experts, and EfficientMoE \cite{zeng2025efficientmoe} predicts device load in real time to deploy hot and cold experts across heterogeneous hardware. These methods primarily treat expert placement and routing as a load-balancing problem, whereas CAP further models the communication cost induced by expert placement.

\paragraph{Expert Pruning} Expert pruning offers another direction for accelerating MoE inference\cite{kim2021scalableefficientmoetraining,yang2024moei2compressingmixtureexperts,lee2025stunstructuredthenunstructuredpruningscalable,zhang2025diversifyingexpertknowledgetaskagnostic,chowdhury2024provablyeffectivemethodpruning}. Not All Experts Are Equal\cite{lu2024expertsequalefficientexpert} estimates expert-level importance via output deviation on a small calibration set and removes low-impact experts without additional training. Task-Specific Expert Pruning\cite{chen2022taskspecificexpertpruningsparse} instead accumulates gating statistics during fine-tuning to identify task-relevant experts and progressively prune unimportant ones. Lynx\cite{gupta2024lynxenablingefficientmoe} further performs dynamic, batch-aware expert reduction at inference time by leveraging router confidence and expert importance hierarchy to drop secondary experts with minimal accuracy loss. In contrast, CAP treats pruning not only as a computation-reduction mechanism but also as a communication optimization problem by explicitly modeling device- and node-level communication cost in pruning decisions.

\clearpage

\bibliographystyle{IEEEtran}
\bibliography{bibfile}

\end{document}